\documentclass[conference]{IEEEtran}
\IEEEoverridecommandlockouts 						
\usepackage{amsmath,amssymb}
\usepackage{amsthm}
\usepackage{color}
\usepackage[mathscr]{eucal}
\usepackage{graphics,graphicx,multicol}
\usepackage{epsfig}
\usepackage{enumerate}
\usepackage{subfigure}
\usepackage{morefloats}
\usepackage{enumerate}
\usepackage{subfigure}
\usepackage{multirow}
\usepackage{array}
\usepackage{pstricks, pst-node, pst-plot, pst-circ}
\usepackage{moredefs}

\usepackage{booktabs}
\usepackage{slashbox}
\usepackage{amsmath}
\usepackage{multirow}
\usepackage{graphicx}
\usepackage{subfig}
\usepackage{algorithm}
\usepackage{algorithmic}
\usepackage{bm}
\ifCLASSINFOpdf
\else
\fi
\usepackage{courier}

\begin{document}
\title{A Utility Proportional Fairness Resource Allocation in Spectrally Radar-Coexistent Cellular Networks}
\author{Mo Ghorbanzadeh, Ahmed Abdelhadi, Charles Clancy\\
Hume Center, Virginia Tech, Arlington, VA, 22203, USA\\
\{mgh, aabdelhadi, tcc\}@vt.edu}
\maketitle

\begin{abstract}
Spectrum sharing is an elegant solution to addressing the scarcity of the bandwidth for wireless communications systems. This research studies the feasibility of sharing the spectrum between sectorized cellular systems and stationary radars interfering with certain sectors of the communications infrastructure. It also explores allocating optimal resources to mobile devices in order to provide with the quality of service for all running applications whilst growing the communications network spectrally coexistent with the radar systems. The rate allocation problem is formulated as two convex optimizations, where the radar-interfering sector assignments are extracted from the portion of the spectrum non-overlapping with the radar operating frequency. Such a double-stage resource allocation procedure inherits the fairness into the rate allocation scheme by first assigning the spectrally radar-overlapping resources.
\end{abstract}
\IEEEpeerreviewmaketitle


\section{Introduction}\label{sec:introduction}
During recent years, mobile subscribers' quantity and their traffic volume have increased so enormously that the monthly wireless traffic has observed a growth of about $7800\%$ and $250\%$ for data and voice respectively \cite{EricssonMobilityReport2013}. In response to such an explosive devour of wireless networks for more bandwidth, the idea of spectrum sharing can staunchly alleviate the spectral scarcity by making communications devices leverage operating frequency bands assigned for other specialty applications such as radars; For instance, the United States National Broadband Plan has aimed at reallocating the S-band radar $3500 - 3650$ MHz spectrum to the broadband wireless access. However, such notions entail severe issues for the operation of spectrum-shared devices.

Since radars are generally high-powered devices as opposed to wireless user equipments (UE)s, they can adversely affect the performance of spectrally coexistent communication systems. Moreover, electromagnetic interference from the communications devices can jeopardise radar missions. Therefore, assigning the radar spectrum for communications operations should refrain from interfering with radars. Since radars are often not in the vicinity of wireless networks, the communication systems can utilize the radar frequencies entirely, and an occasional radar proximity to the wireless network should be incorporated into spectrum sharing designs by which cellular networks do not operate in radar bands.

On the other hand, the prevalence of smartphones running applications with distinct quality of service (QoS) requirements \cite{GhorbanzadehICNC2013} makes efficient an bandwidth allocation indispensable. Because applications QoS walks hand in hand with efficient resource allocation, any effort toward spectrum-shared rate allocation should be realized with QoS in mind. As such, in this paper, we will apply algorithms to allocate the radar spectrum optimally to sectorized cellular networks abstaining from any sector-radar interference, where the allocation is formulated as convex optimization problems.

Next, section \ref{sec:related} surveys the topical literature concerning radar-coexistent communications systems and their resource allocations.

\section{Related Work}\label{sec:related}
Radar coexistent communications systems have not received their deserved research attention so far. The authors of \cite{Lackpour:2011} investigated WiMax-radar spectral and temporal interference mitigation techniques. Multiple carriers increasing of the available bandwidth for smart phones was introduced in \cite{Yuan2010}. The authors of \cite{kelly98ratecontrol} presented a proportional fairness resource allocation optimization for communications networks using lagrange multipliers. \cite{Shajaiah:Milcom:2013} proposed an optimal rate allocation for dual carrier systems containing real-time and delay-tolerant applications. In \cite{Lee:2005}, the authors suggested a distributed power allocation for cellular systems with sigmoidal utility functions to approximate a global optimal solution at the expense of dropped-users by which it could not guarantee a minimum QoS level for the UEs. In \cite{Shajaiah:ICNC:2013}, the authors presented an optimal rate allocation algorithm for users of a single carrier. They 
formulated the rate allocation in a convex framework including logarithmic and sigmoidal utility functions as delay-tolerant and real-time applications respectively. The authors of \cite{Ahmed:ICNC:2014} devised a rate allocation algorithm prioritized real-time applications over delay-tolerant ones by using a utility proportional fairness policy. In another work, \cite{Ahmed:PIMRC:2013} proposed a rate assignment for the LTE networks with elastic/inelastic traffic types, represented as logarithmic/sigmoidal utility functions, that avoided rate fluctuations in peak-traffic hours. In \cite{Kurrle2012}, the author presented a weighted aggregation of elastic and inelastic traffic utility functions approximated to the nearest concave utility from a set of functions through the minimum mean squared error measure. The approximate utility was used to solve the rate allocation problem through a modification of the distributed algorithm introduced in \cite{kelly98ratecontrol} such that the solution approximated the 
optimal rates. The authors in \cite{Tao2008} proposed a subcarrier allocation for multiuser orthogonal frequency division multiplexing systems which paid attention to the delay-sensitive traffic and used network delay models, which can be inferred \cite{Ghorbanzadeh2013}, to assign the subcarriers. Finally, \cite{Ahmed:Dyspan:2014} proposed a resource allocation architecture which considers context, time, and location to assign resources to applications.

Our contributions in this paper are: I) We introduce an optimal rate allocation for radar-coexistent cellular networks with interference-refraining sectors. II) We demonstrate that the interfering sectors rates change as a radar approaches the cellular infrastructure. III) We elucidate that the proposed method fairly allocates resources even when a radar operates close-by. The remainder of this paper proposes the problem formulation for resource allocation problem in section \ref{sec:SysModel}, leverages algorithms to solve the problem in section \ref{sec:algorithm}, sets up simulations and discusses relevant quantitative results in section \ref{Section:sim}, and concludes the paper in  section \ref{Sec:Conclusion}.

\begin{figure}[!htb]
\begin{center}
\includegraphics[width=3.5in]{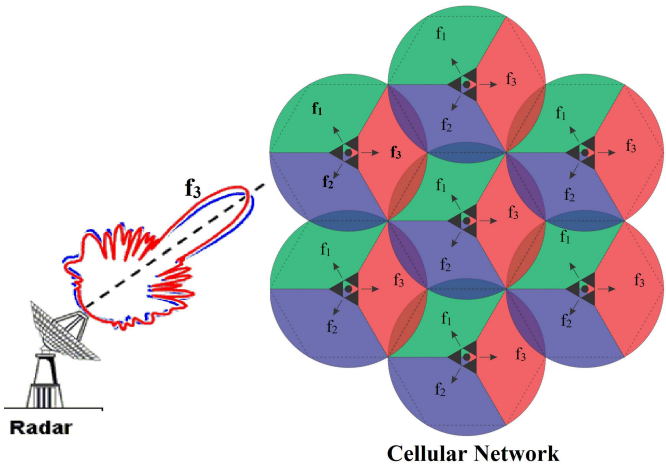}
\end{center}
\caption{\footnotesize{Cellular communications system: Cell sector colors indicate frequency bands ($f_1$, $f_2$, $f_3$) and identically colored sectors imply frequency reuse whose topological pattern minimizes the inter-cell interference. The radar and pink sectors operations in the same frequency band $f_3$ create interferes.}} \label{fig:CellularInter}
\end{figure}

\section{System Model} \label{sec:SysModel}
Envisage a hexagonal $K$-cell $L$-sector communications network with the frequency reuse and a per-cell base station denoted by the long term evolution (LTE) nomenclature as evolved node-B (eNB), controlled by the Mobile Management Entity (MME). A radar operation close by the cellular system causes interference with sectors working at the radar frequency. We assume a relatively stationary radar interferes with deterministic sectors inasmuch as topologically identical sectors deploy the frequency reuse.For the exemplar network in Fig. \ref{fig:CellularInter} with the eNBs shown as the gray triangular shapes at the circle centers, the $f_3$ Hz-operating pink sectors and radar can interfere while the green and pink sectors work safely in different frequency bands ($f_1$,$f_2$) belonging solely with the cellular network. It is worth mentioning that such a sector frequency pattern reduces the inter-cell interference by spatially maximizing co-channels, i.e. identical frequency sectors.

\begin{figure}[!htb]
\begin{center}
\includegraphics[width=3.5in]{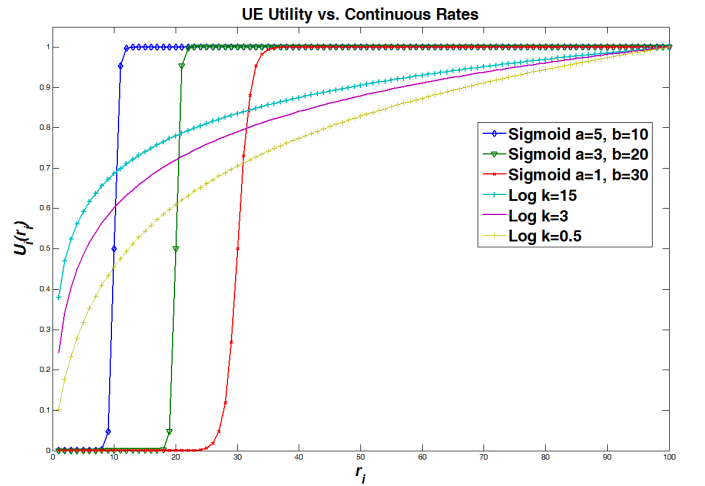}
\end{center}
\caption{Three sigmoidal (purple, green, red) and logarithmic ( yellow, pink, cyan) utilitiy functions  for real-time and delay tolerant applications.} \label{fig:AppUtlFn}
\end{figure}

Moreover, the cellular system includes $M$ UEs running a delay-tolerant or real-time application mathematically representable by sigmoidal and logarithmic utility functions, the user's service satisfaction vs. the UE rate, as in the equations (\ref{eqn:sigmoid}) and (\ref{eqn:log}) respectively, which are nonnegative, strictly increasing, continuous, and zero-valued at zero rates \cite{Tychogiorgos:PIMRC:2012}.

\begin{equation}\label{eqn:sigmoid}
U_i(r_i) = c_i\Big(\frac{1}{1+e^{-a_i(r_i-b_i)}}-d_i\Big)
\end{equation}

\begin{equation}\label{eqn:log}
U_i(r_i) = \frac{\log(1+k_ir_i)}{\log(1+k_i r_{max})}
\end{equation}

where $c_i = \frac{1+e^{a_ib_i}}{e^{a_ib_i}}$, $d_i = \frac{1}{1+e^{a_ib_i}}$, $r_{max}$ is the $100\%$ satisfaction-achieving rate ($U(r_{max})=1$), and $k_{i}$ is the utility increase with enlarging the $r_{i}$. Based on \cite{Ahmed:ICNC:2014}, $r_i = b_i$ is the inflection point of the sigmoidal utility function (\ref{eqn:sigmoid}) such that only for $r_i > b_i$ the user is satisfied. Parameters $(a_i, b_i, c_i, k_{i}, d_i)$ direct impact on the utility shape can model various applications such as VoIP ($a_i=5$ and $b_i=10$), video streaming ($a_i=0.5$ and $b_i=20$), and FTP ($k=15$) \cite{Ahmed:ICNC:2014}. Six sample utility functions are depicted in Fig. \ref{fig:AppUtlFn}.

The $i^{th}$ UE rate allocated by its eNB $l^{th}$ sector, where $l \in \{1,...,L\}$, includes constituents from radars spectrum, $r_{i,\text{radar}}^{l}$, and communications spectrum, $r_{i,\text{comm}}^{l}$, so that $r_{i,ag}^{l} = r_{i,\text{radar}}^{l} + r_{i,\text{comm}}^{l}$. Therefore, the $i^{th}$ UE utility, a function of its aggregate rate assigned by all sectors, can be expressed as in the equation (\ref{eqn:utagrt}) below.

\begin{equation}\label{eqn:utagrt}
U_i(r_{i,ag}) = U_i(\sum_{l=1}^{L} r_{i,ag}^{l}) = U_i(\sum_{l=1}^{L}(r_{i,\text{comm}}^{l} + r_{i,\text{comm}}^{l})).
\end{equation}

Next, section \ref{sec:rsal} develops the mathematical formulation for the rate allocation optimization.

\subsection{Resource Allocation}\label{sec:rsal}
To guarantee a minimum service QoS, a minimal nonzero rate per UE is realized through a proportional fairness (no user drop occurs) \cite{Ahmed:ICNC:2014}. Moreover, spectral coexistence with nearby radars resembles the dual carrier rate allocation optimization in \cite{Shajaiah:Milcom:2013}, where UEs gain escalated rates by leveraging available bandwidths simultaneously from primary and secondary carriers, where the former and latter are respectively the radar and cellular systems in our proposed system model. Although a primary carrier is conventionally the communications side in the literature, section \ref{Section:sim} shows this situation lacks fairness of rate assignments in that radar spectrum-coinciding frequencies can only be allotted to sectors which do not interfere with the radar operation. We propose a two-stage resource allocation, whose stage 1 assigns resources from the radar spectrum as in the equation (\ref{eqn:pr_fairness}) where $r_{i,\text{radar}}$ is the radar spectrum allocated to the 
$i^{th}$ UE, $R_{\text{radar}}^{l}$ is the stage 1 maximum achievable rate allocated by the MME to the cells $l^{th}$ sectors, $M_{k}$ is the $k^{th}$ cell UE quantity, $\textbf{r}_{\textbf{\text{radar}}} = \{r_{1,\text{radar}}, \ldots, r_{M,\text{radar}}\}$, and $R_{\text{radar}}^{l_{\text{interference}}} = 0$ implies that no rate is allocated to radar-interfering sectors at this stage which excerpts resources from the radar spectrum.

\begin{equation}\label{eqn:pr_fairness}
\begin{aligned}
& \underset{\textbf{r}_{\textbf{radar}}}{\text{max}}
& & \prod_{i=1}^{M}U_i(\sum_{l=1}^{L}r_{i,\text{radar}}^{l}) \\
& \text{subject to}
& & \sum_{l=1}^{L}r_{i,\text{radar}}^{l} = r_{i,\text{radar}}, \sum_{l=1}^{M_{k}}r_{i,\text{radar}}^{l} \leq R_{\text{radar}}^{l}\\ & & &\sum_{l=1}^{L}R_{\text{radar}}^{l} = R_{\text{radar}}, R_{\text{radar}}^{l_{\text{interference}}} = 0, r_{i,\text{radar}}^{l} \geq 0\\
& & & i = 1,...,M, l = 1,...,L.
\end{aligned}
\end{equation}

Next, at stage 2 the carrier (communications) spectrum is allocated to the UEs based on the equation (\ref{eqn:secon_fairness}) where $r_{i,\text{comm}}$ is the $i^{th}$ UE allocated communications spectrum, $R_{\text{comm}}^{l}$ is the $l^{th}$ sector maximum achievable rate once resources are excerpted from the carrier spectrum (not radar), $\textbf{r}_{\textbf{comm}} = \{r_{1,\text{comm}}, \ldots, r_{M,\text{comm}}\}$, and $R_{\text{comm}}^{l}$ is the communications spectrum allocated to the $l^{th}$ sector of all the cells by the MME. It is notable that shifting non-interfering cell rates by the stage 1-assigned amounts is included in the equation (\ref{eqn:secon_fairness})to insure fairness (we will see in section \ref{Section:sim}).

\begin{equation}\label{eqn:secon_fairness}
\begin{aligned}
& \underset{\textbf{r}_{\textbf{comm}}}{\text{max}}
& & \prod_{i=1}^{M}U_i(r_{i,\text{comm}} + r_{i,\text{radar}}^{\text{opt}}) \\
& \text{subject to}
& & \sum_{l=1}^{L}r_{i,\text{comm}}^{l} = r_{i,\text{comm}}, \sum_{l=1}^{M_{k}}r_{i,\text{comm}}^{l} \leq R_{\text{comm}}^{l}\\
& & & \sum_{l=1}^{L}R_{\text{comm}}^{l} = R_{\text{comm}}, r_{i,\text{radar}}^{\text{opt}} = \sum_{l = 1}^{L}r_{i,\text{radar}}^{l,\text{opt}},\\ 
& & & r_{i,\text{comm}}^{l} \geq 0, i = 1,...,M, l = 1,...,L.
\end{aligned}
\end{equation}

It is proved that a resource allocation problem in the form of the equations (\ref{eqn:pr_fairness}) and (\ref{eqn:secon_fairness}) is a convex optimization \cite{Ahmed:PIMRC:2013} which means their global maxima are tractable respectively as $\{r_{i,\text{radar}}^{l,\text{opt}} | i \in \{1,\ldots,M\} \wedge l \in \{1,\ldots,L\}\}$ and $\{r_{i,\text{comm}}^{l,\text{opt}} | i \in \{1,\ldots,M\} \wedge l \in \{1,\ldots,L\}\}$, which are the set of optimal rates allocated to the $i^{th}$ UE by $l^{th}$ sector from the radar and communications spectrum, in that order. The UE total optimal rate in the system is sum of the aforementioned solutions. To summarize, for the $i^{th}$ UE, we first get $r_{i,\text{radar}}^{l,\text{opt}}$ (which is $0$ radar-interfering UEs), then we obtain $r_{i,\text{comm}}^{l,\text{opt}}$, and the total optimal rate will be $r_{i,ag}^{l,\text{opt}} = r_{i,\text{comm}}^{l,\text{opt}} + r_{i,\text{radar}}^{l,\text{opt}}$.

Next, section \ref{sec:algorithm} presents suitable algorithms to solve the equations (\ref{eqn:pr_fairness}) and (\ref{eqn:secon_fairness}).

\section{Algorithm}\label{sec:algorithm}
Equation (\ref{eqn:pr_fairness})/(\ref{eqn:secon_fairness}) solution relies on the Algorithms (\ref{alg:UE})/(\ref{alg:UE2}), (\ref{alg:eNodeB}), and (\ref{alg:MME}), of which Algorithm (\ref{alg:MME}) is a variation of the algorithm in \cite{Ahmed:Dyspan:2014} and the others are modifications of the dual carrier resource allocation algorithms in \cite{Shajaiah:Milcom:2013}. The subscript "radar"/"comm" in the Algorithm (\ref{alg:UE})/(\ref{alg:UE2}) indicates resource allocations from radars/communications spectrum, which replaces the general subscript $j$ in the Algorithms (\ref{alg:eNodeB}) and (\ref{alg:MME}) at stage 1/stage 2 of the resource allocation to imply that resources are assigned from the radar/communications spectrum only. To recap, at stage 1 we allocate from the radar spectrum to non-interfering sectors of the cells and no resources are passed to radar-interfering sectors' UEs. To ensure the fairness, less new resources should be given to the non-interfering sectors' UEs at stage 2 as radar-
interfering UEs did not yet receive any stage 1 bandwidth. This is realized by the shift in the Algorithm (\ref{alg:UE2}) with the subscript "radar" incorporating rate shifts due to the stage 1 assignments from the radar spectrum.

\begin{algorithm}
\caption{The $l^{th}$ Sector $i^{th}$ UE Algorithm - \textbf{Stage 1}, \cite{Shajaiah:Milcom:2013}}\label{alg:UE}
\begin{algorithmic}
\STATE {Send an initial bid $w_{i}^{l}(1)$ to the eNB $l^{th}$ sector.}
\LOOP
	\STATE {Receive a shadow price $P_{l}(n)$ from the eNB $l^{th}$ sector.}
	\IF {STOP from the eNB $l^{th}$ sector,}
    \STATE {Calculate the allocated rate $r_{i,\text{radar}}^{l,\text{opt}} = \frac{w_{i,\text{radar}}^{l}(n)}{P_{\text{radar}}^{l}(n)}$.}
	\ELSE
    \STATE {Calculate $r_{i,\text{radar}}^{l}(n) = \arg \underset{r_{i,\text{radar}}^{l}}\max \Big(\log U_i(r_{i,\text{radar}}^{l}) - P_{\text{radar}}^{l}(n)r_{i,\text{radar}}^{l}\Big)$.}
	\STATE {Calculate a new bid $w_{i,\text{radar}}^{l} (n)= P_{\text{radar}}^{l}(n) r_{i,\text{radar}}^{l}(n)$.}
    \STATE {Send the new bid $w_{i,\text{radar}}^{l} (n)$ to the eNB $l^{th}$ sector.}
	\ENDIF
\ENDLOOP
\end{algorithmic}
\end{algorithm}

\begin{algorithm}
\caption{The $l^{th}$ Sector $i^{th}$ UE Algorithm - \textbf{Stage 2}, \cite{Shajaiah:Milcom:2013}}\label{alg:UE2}
\begin{algorithmic}
\STATE {Send an initial bid $w_{i,\text{comm}}^{l}(1)$ to the eNB $l^{th}$ sector.}
\LOOP
	\STATE {Receive a shadow price $P_{\text{comm}}^{l}(n)$ from the eNB $l^{th}$ sector.}
	\IF {STOP from the eNB $l^{th}$ sector.}
    \STATE {Calculate the allocated rate $r_{i,\text{radar}}^{l,\text{opt}} = \frac{w_{i,\text{radar}}^{l}(n)}{P_{\text{radar}}^{l}(n)}$.}
	\ELSE
    \STATE {Calculate $r_{i,\text{comm}}^{l}(n) = \arg \underset{r_{i,\text{comm}}^{l}}\max \Big(\log U_i(r_{i,\text{comm}}^{l} + r_{i,\text{radar}}^{l,\text{opt}}) - P_{\text{comm}}^{l}(n)r_{i,\text{comm}}^{l}\Big)$.}
	\STATE {Calculate a new bid $w_{i,\text{comm}}^{l} (n)= P_{\text{comm}}^{l}(n) r_{i,\text{comm}}^{l}(n)$.}
    \STATE {Send the new bid $w_{i,\text{comm}}^{l} (n)$ to the eNB $l^{th}$ sector.}
	\ENDIF
\ENDLOOP
\end{algorithmic}
\end{algorithm}

\begin{algorithm}
\caption{The eNB $l^{th}$ Sector Algorithm - \textbf{Stage 1,2}, \cite{Shajaiah:Milcom:2013}}\label{alg:eNodeB}
\begin{algorithmic}
\LOOP
	\STATE {Receive UE bids $w_{i,j}^{l}(n)$.}
    \STATE {Calculate aggregate bids $W_{k,j}^{l}(n)$ and send them to the MME.}
	\COMMENT{Let $w_{i,j}^{l}(0) = 0\:\:\forall i$}
    \STATE {Receive the sector rate $R^{l}(n)$ from the MME.}
    \IF	{STOP received from the MME,}
    \STATE {STOP and send STOP to all UEs.}
    \ELSE
    \STATE {Calculate $P_{j}^{l}(n) = \frac{\sum_{i = 1}^{M}w_{i,j}^{l}(n)}{R_{j}^{l}}$.}
    \STATE {Send the new shadow price to all UEs.}
	\ENDIF
\ENDLOOP
\end{algorithmic}
\end{algorithm}

\begin{algorithm}
\caption{The MME Algorithm - - \textbf{Stage 1,2}, \cite{Ahmed:Dyspan:2014}}\label{alg:MME}
\begin{algorithmic}
\STATE {Send the sector rate $R^{l}(0)$ to the $l^{th}$ sector.}
\COMMENT {Let $R^{l}(0) = \frac{R}{L}$.}
\LOOP
	\STATE {Receive aggregated bids from $W_{k,j}^{l}(n)$ from the $l^{th}$ sector.}
    \STATE {Calculate total aggregated bids $W_{j}^{l}(n)$.}
	\COMMENT{Let $W_{j}^{l}(0) = 0\:\:\forall l$}
    \STATE {Receive a sector rate $R^{l}(n)$ from the MME.}
    \IF	{$|W_{j}^{l}(n) - W_{j}^{l}(n-1)| < \delta  \:\:\forall l$,}
    \STATE {STOP and send STOP to all sectors.}
    \ELSE
    \STATE {Calculate $R_{j}^{l}(n)$ and send to the $l^{th}$ sector.}
	\ENDIF
\ENDLOOP
\end{algorithmic}
\end{algorithm}

The resource allocation process is as follows. The $i^{th}$ UE sends its initial bid $w_{i,j}^{l}$ to the $k^{th}$ eNB $l{th}$ sector, which calculates the aggregate sector bid $W_{k,\text{radar}}^{l}(n)$ at time $n$ and transmits the aggregate sector bids $\{W_{k,\text{radar}}^{l}(n) | l = \{1, \ldots, L\}\}$ to the MME. This entity computes total aggregate sector bids $W_{\text{radar}}^{l}$ and the difference from its former value $|W_{\text{radar}}^{l}- W_{\text{radar}}^{l-1}|$ for all the sectors. Should the difference be less than a pre-set threshold $\delta$ for all the sectors, an \emph{exit} criterion is met; Otherwise, the MME evaluates sector rates $R_{\text{radar}}^{l}$ and transmit them to corresponding eNBs. Furthermore, the $k^{th}$ eNB $l{th}$ sector calculates the shadow price (price per unit bandwidth for all communications channels) $P_{radar}^{l}(n) = \frac{\sum_{i = 1}^{M}w_{i,\text{radar}}^{l}(n)}{R_{\text{radar}}^{l}s}$ and transmits it to its covered UEs which solve $r_{i,\text{radar}}
^{l}(n) = \arg \underset{r_{i,\text{radar}}^{l}}\max \Big(\log U_i(r_{i,\text{radar}}^{l} - P_{\text{radar}}^{l}(n)r_{i,\text{radar}}^{l})\Big)$. Then, the calculated bid $w_{i,radar}^{l}(n) = P_{\text{radar}}^{l}(n)r_{i,\text{radar}}^{l}(n)$ is sent to the $l{th}$ sector eNB and the procedure is repeated until the termination of the stage 2, $|W_{\text{radar}}^{l}- W_{\text{radar}}^{l-1}| < \delta$. The final optimum rates are calculated as $r_{i,\text{radar}}^{l,\text{opt}} = \frac{w_{i,\text{radar}}^{l}(n)}{P_{\text{radar}}^{l}(n)}$. 

\begin{figure}[!htb]
\begin{center}
\includegraphics[width=3.5in]{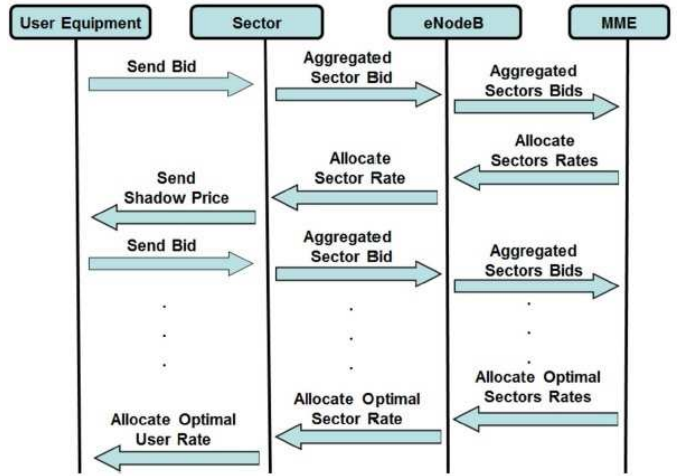}
\end{center}
\caption{\footnotesize{Resource Allocation Stage Diagram: This illustration only depicts one stage of the allocation whereas the assignment is performed once over the radar spectrum (stage 1) then over the carrier spectrum where radar is not operating (stage 2). The procedure in both stages is similar with the difference that the stage 2 includes a rate shift to include the stage 1-allotted bandwidth from radars spectrum (adopted from \cite{Ahmed:Dyspan:2014}).}} \label{fig:stagediagram}
\end{figure}

Then, at stage 2, each UE transmits an initial bid $w_{i,\text{comm}}^{l}(1)$ to the $k^{th}$ eNB $l{th}$ sector, which calculates the aggregate sector bids $W_{k,\text{comm}}^{l}(n)$ at time $n$ and transmits it to the MME, which, in case $|W_{\text{comm}}^{l}- W_{\text{comm}}^{l-1}| > \delta$, evaluates sector rates $R_{j}^{l}$ and sends them to corresponding eNBs, which transmit the UEs their calculated shadow price $P_{\text{comm}}^{l}(n) = \frac{\sum_{i = 1}^{M}w_{i,\text{comm}}^{l}(n)}{R_{\text{comm}}^{l}}$ leveraged to evaluate UE rates $r_{i,\text{comm}}^{l}(n) = \arg \underset{r_{i,\text{comm}}^{l}}\max \Big(\log U_i(r_{i,\text{comm}}^{l} + r_{i,\text{radar}}^{l,\text{opt}}) - P_{\text{comm}}^{l}(n)r_{i,\text{comm}}^{l}\Big)$. It is noteworthy that $r_{i,p}^{l,\text{opt}}$ in the Algorithm \ref{alg:UE2} is  stage 1-obtained rate (solution of the Algorithm \ref{alg:UE}) and is part of the radar operating frequency. Then, the calculated bid $w_{i,comm}^{l}(n) = P_{\text{comm}}^{l}(n)r_{i,\text{comm}}^{
l}(n)$ is sent to the $l{th}$ sector eNB and the procedure is repeated until the termination of the stage 2, $|W_{\text{comm}}^{l}- W_{\text{comm}}^{l-1}| < \delta$. The final optimum rates are calculated as $r_{i,\text{comm}}^{l,\text{opt}} = \frac{w_{i,\text{comm}}^{l}(n)}{P_{\text{comm}}^{l}(n)}$. This entire process is summarized in Fig. \ref{fig:stagediagram}.

Next section simulates the resource allocation modus operandi of the section \ref{sec:algorithm} for a cellular system, with various UE applications, spectrally coexistent with a radar.

\section{Simulations} \label{Section:sim}
This section starts with simulating delay tolerant and real-time applications based on the sigmoidal and logarithmic utility functions in our system. Then, it implements the algorithms presented in the section \ref{sec:algorithm} for a cellular network in the vicinity of a spectrally coexistent radar. All simulations are performed in MATLAB. 

\begin{figure}[!htb]
\begin{center}
\includegraphics[width=3.5in]{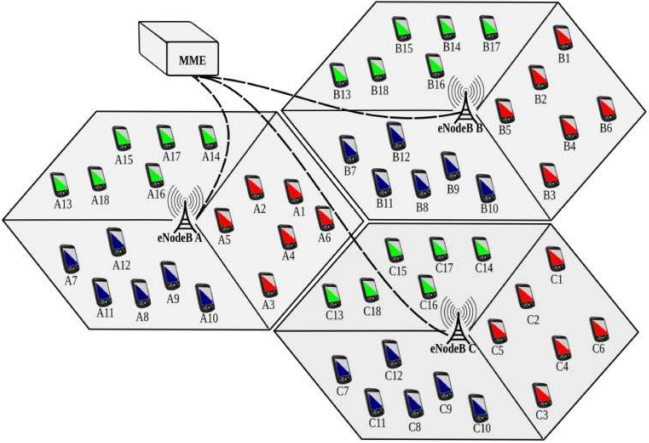}
\end{center}
\caption{\footnotesize{Cellular communications system: Three sector cells are covered by eNBs, controlled by an MME unit. The cell phone callers indicate frequency bands such that sectors with the same colored cell phones are reusing identical bands. Such a topological reuse pattern is aimed at minimizing the inter-cell interference.}} \label{fig:CellularCom}
\end{figure}

We consider a cellular network as Fig. \ref{fig:CellularCom} with 3 sectors, operating in different frequency bands, whose topologically identical sectors form co-channels. For instance, the blue, red, and green cell phones indicate that their host sectors operate in distinct frequencies repeated with the same pattern over other cells of the infrastructure. Each cell is equipped with an eNB, in charge of the UEs under its coverage area, collaborating with an MME unit which monitors the operation of all the eNBs in the cellular network. The UEs run different applications whose utility function parameters are shown in Table \ref{table:Param}, which represent sectors of the cells A, B, and C of the network in Fig. \ref{fig:CellularCom} and 6 applications inside each sector. The 3 sigmoidal/logarithmic application utilities per sector are characterized with the abbreviation "Sig"/"Log". For example, Table \ref{table:Param} says that the UE C5 is in cell C sector 1 and runs a logarithmic (delay-tolerant) 
application with parameter $k=1.8$. Furthermore, we assume an approaching radar and red UE sectors use the same spectrum and 200/400 bandwidth units are available to the radar/communications system at maximum.

\begin{table}[h]
\caption{Application Utility Functions of the Cellular Network.}
\centering
\begin{tabular}{c c c c | c c c}
\hline\hline
\textbf{Sector 1 - Cell A}\\[0.5ex]
\hline
A1 & Sig & a = 3 & b = 10.0 & A4 & Log & k = 1.1 \\
A2 & Sig & a = 3 & b = 10.3 & A5 & Log & k = 1.2 \\
A3 & Sig & a = 1 & b = 10.6 & A6 & Log & k = 1.3\\
\hline
\textbf{Sector 2 - Cell A}\\[0.5ex]
\hline
A7 & Sig & a = 3 & b = 10.0 & A10 & Log & k = 1 \\
A8 & Sig & a = 3 & b = 15.3 & A1  & Log & k = 2 \\
A9 & Sig & a = 3 & b = 12.0 & A12 & Log & k = 3\\
\hline
\textbf{Sector 3 - Cell A}\\[0.5ex]
\hline
A13 & Sig & a = 3 & b = 15.1 & A16 & Log & k = 10 \\
A14 & Sig & a = 3 & b = 15.3 & A17 & Log & k = 11 \\
A15 & Sig & a = 3 & b = 15.5 & A18 & Log & k = 12\\
\hline
\textbf{Sector 1 - Cell B}\\[0.5ex]
\hline
B1 & Sig & a = 3 & b = 15.9 & B4 & Log & k = 1.4 \\
B2 & Sig & a = 3 & b = 11.2 & B5 & Log & k = 1.5 \\
B3 & Sig & a = 1 & b = 11.5 & B6 & Log & k = 1.6\\
\hline
\textbf{Sector 2 - Cell B}\\[0.5ex]
\hline
B7 & Sig & a = 3 & b = 13 & B10 & Log & k = 4 \\
B8 & Sig & a = 3 & b = 14 & B11 & Log & k = 5 \\
B9 & Sig & a = 1 & b = 15 & B12 & Log & k = 6\\
\hline
\textbf{Sector 3 - Cell B}\\[0.5ex]
\hline
B13 & Sig & a = 3 & b = 15.7 & B16 & Log & k = 13 \\
B14 & Sig & a = 3 & b = 15.9 & B17 & Log & k = 14 \\
B15 & Sig & a = 3 & b = 17.3 & B18 & Log & k = 15\\
\hline
\textbf{Sector 1 - Cell C}\\[0.5ex]
\hline
C1 & Sig & a = 3 & b = 11.8 & C4 & Log & k = 1.7 \\
C2 & Sig & a = 3 & b = 12.1 & C5 & Log & k = 1.8 \\
C3 & Sig & a = 1 & b = 12.4 & C6 & Log & k = 1.9\\
\hline
\textbf{Sector 2 - Cell C}\\[0.5ex]
\hline
C7 & Sig & a = 3 & b = 16 & C10 & Log & k = 7 \\
C8 & Sig & a = 3 & b = 17 & C11 & Log & k = 8 \\
C9 & Sig & a = 1 & b = 18 & C12 & Log & k = 9\\
\hline
\textbf{Sector 3 - Cell C}\\[0.5ex]
\hline
C13 & Sig & a = 3 & b = 17.5 & C16 & Log & k = 16 \\
C14 & Sig & a = 3 & b = 17.7 & C17 & Log & k = 17 \\
C15 & Sig & a = 3 & b = 17.9 & C18 & Log & k = 18\\
\hline
\end{tabular}
\label{table:Param}
\end{table}

\begin{figure}[!htb]
\begin{center}
\includegraphics[width=3.5in]{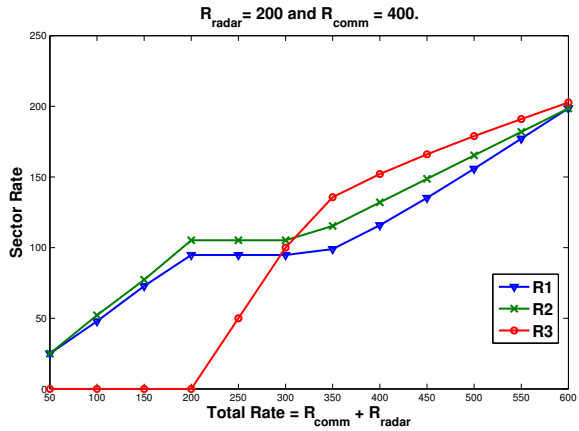}
\end{center}
\caption{\footnotesize{Rate allocation to the UEs when a radar interfering with sectors 3 of cells is in the vicinity of the cellular communications system. The allocated rates are initially allocated to non-interfering sectors from the radar operating frequency bands and no bandwidth is allocated to the interfering sectors. In the second stage of the allocation, the rates start allocating to the interfering sectors only in order to insure fairness. Once certain bandwidth is allocated to the sector, the remainder of the bandwidths can be allocated to all the sectors.}} \label{fig:AllocationShared}
\end{figure}

The sector rates are plotted in Fig. \ref{fig:AllocationShared}, which shows the more resources available to eNBs, the higher rate the sector rates. Furthermore, the radar-interfering red sector rates are $0$ at stage 1 (Algorithms (\ref{alg:UE}), (\ref{alg:eNodeB}), and (\ref{alg:MME})), whilst the non-interfering blue and green sectors are assigned resources from the radar spectrum. At stage 2 (Algorithms (\ref{alg:UE2}, (\ref{alg:eNodeB}), and (\ref{alg:MME})) non-radar carrier spectrum is allocated to all sectors. Particularly, the sharp increase of the red sectors rate vs. the other ones is due to no resource assignments to the red sectors at the preceding stage 1. At this time, no $R_{\text{comm}}^{1}$ and $R_{\text{comm}}^{2}$ allocations occur until $R_{\text{comm}}^{3}$ equals their rates after which these non-interfering sectors also obtain more resources; Hence, giving the radar spectrum to non-interfering sectors initially then assigning much more radar-spectrum resources to interfering sectors 
hold an intrinsic fairness into the rate assignment procedure. Once some resources are passed to the interfering red sectors, the rate plots grow very close to each other due to the traffic analogy in the sectors (3 sigmoidal and 3 logarithmic utility).

\begin{figure}[!htb]
\begin{center}
\includegraphics[width=3.5in]{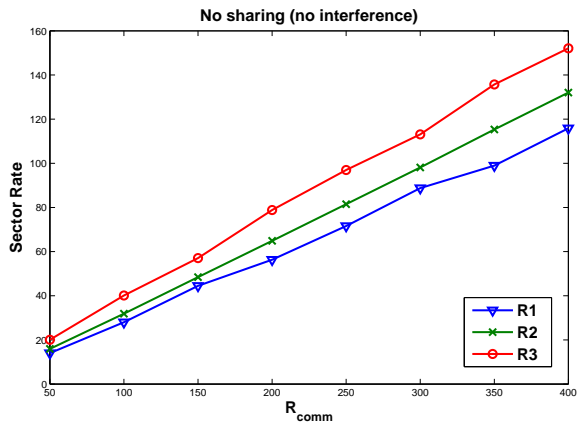}
\end{center}
\caption{\footnotesize{Rate allocation to the UEs when no radar is in the vicinity of the cellular communications system. The allocated rates are similar since the pattern of the applications in the cells are alike.}} \label{fig:AllocationNotShared}
\end{figure}

To compare the deployment of the devised resource allocation and lack of it, we do the same experiment under the assumption of no shared spectrum, for which rate allocations are plotted in Fig. \ref{fig:AllocationNotShared}. As we see, the red sector rates are allocated from the commencement of the allocation since no interfering sector exists. Besides, Fig. \ref{fig:AllocationShared} shows that, at the rate $250$, $R_{\text{comm}}^{3}$ utilizes $50$ bandwidth units from the communications spectrum whereas, before the rate $200$, its allocated rate is $0$. On the flip side, \ref{fig:AllocationNotShared} depicts that at the rate $250$, $R_{\text{comm}}^{3}$ uses much more units of bandwidth.

\section{Conclusion}\label{Sec:Conclusion}
In this paper, we presented a two-stage novel resource allocation optimization method which assigned bandwidth to the UEs of cellular communications networks operating in the vicinity of a radar such that the radar spectrum was shared with certain sectors of the communications infrastructure. First, we formulated the rate allocation process as two convex optimization problems which initially assigned resources from the radar spectrum to the UEs in the non-interfering sectors. Then, we allocated the communications spectrum , not coinciding with that of the radar, to all the sectors. The final solution was a set of optimal rates for the UEs in the cellular environment based on their running applications. Moreover, we discussed that, intrinsic to the proportional fairness formulation, the rate allocations dropped no users so that it could warrant a minimum QoS for all running applications.

We demonstrated that the devised two-stage resource allocation scheme not only refrained from any interference between the radar system and sectors of the cellular infrastructure, but it also rendered the rate allocation mechanism fair by incorporating the amount of radar spectrum assigned to non-interfering sectors during the first stage of the allocation into the resource allocation optimization problem at the second stage. Thereby, the interfering sectors were given more resources at the commencement of the second stage to compensate for the lack of assignment in the first stage. Our simulation results validated that the proposed modus operandi afforded the cellular communications network to allocate resources to UEs both to provision applications QoS and to eschew from any interference with neighboring spectrally coexistent radar systems, simultaneously.

\bibliographystyle{ieeetr}
\bibliography{pubs,ref_cite}
\end{document}